\documentclass[journal]{IEEEtran}

\usepackage{graphicx}
\usepackage{xspace}
\usepackage{mathrsfs}

\usepackage{multirow}
\usepackage{makecell}
\usepackage{hhline}
\usepackage{colortbl}

\usepackage{hyperref}
\usepackage{amsmath}
\usepackage{cite}
\usepackage{threeparttable}

\newcommand{\pstate}{$\vec{p}_\text{state}$\xspace}
\newcommand{\partif}{$p_\text{artifact}$\xspace}
\newcommand{\fscore}{$F_1$~score\xspace}
\newcommand{\fscores}{$F_1$~scores\xspace}
\newcommand{\bfscore}{$\bar{F}_1$~score\xspace}
\newcommand{\bfscores}{$\bar{F}_1$~scores\xspace}

\newcommand{\precision}{\ensuremath{p_\text{precision}}\xspace}
\newcommand{\recall}{\ensuremath{p_\text{recall}}\xspace}

\newcommand{\citep}[1]{\cite{#1}}

\begin{document}

\title{Automated Classification of Sleep Stages and EEG Artifacts in Mice with Deep Learning}

\author{Justus T.~C.~Schwabedal, Daniel~Sippel, Moritz~D.~Brandt, Stephan~Bialonski
\thanks{Justus T.~C. Schwabedal, Department of Biomedical Informatics, Emory University, GA, USA --- Daniel Sippel,  Department of Psychiatry and Psychotherapy, University Hospital T\"ubingen, T\"ubingen, Germany; Institute of Medical Psychology and Behavioral Neurobiology, University of T\"ubingen, T\"ubingen, Germany --- Moritz D. Brandt, Department of Neurology, Technische Universit\"at Dresden, Dresden, Germany; German Center for Neurodegenerative Diseases (DZNE) Dresden, Dresden, Germany --- Stephan Bialonski, FH Aachen University of Applied Sciences, Aachen, Germany.}
\thanks{Manuscript received XXX, revised XXX}}

\maketitle

\begin{abstract} 
Sleep scoring is a necessary and time-consuming task in sleep studies.
In animal models (such as mice) or in humans, automating this tedious process
promises to facilitate long-term studies and to promote sleep biology as a data-driven field.  We introduce a deep neural network model that is able
to predict different states of consciousness (Wake, Non-REM, REM) in mice from EEG and EMG recordings
with excellent scoring results for out-of-sample data.  Predictions are made on epochs 
of 4 seconds length, and epochs are classified as artifact-free or not. The model architecture
draws on recent advances in deep learning and in convolutional neural networks research.
In contrast to previous approaches towards automated sleep scoring, our model does not
rely on manually defined features of the data but learns predictive features automatically.
We expect deep learning models like ours to become widely applied in different fields,
automating many repetitive cognitive tasks that were previously difficult to tackle.
\end{abstract}


\section{Introduction}
Humans and other animals spend much of their time in sleep.  Despite its importance, many aspects of sleep are not yet fully understood. In order to uncover mechanisms and functions of sleep, scientists study humans and animal models. Rodents (mice and rats) are often studied as these are well established mammal models which show sleep characteristics that are comparable to human sleep (i.e., NREM and REM sleep). An important step in many studies is the identification of different states of consciousness from different recording modalities such as electroencephalography (EEG) and electromyography (EMG). For many years, sleep scoring has been a manual and time consuming process. The time required for manual sleep scoring depends on various factors (i.e., data quality, artifacts, sleep fragmentation). While experienced personnel needs up to 3 hours to thoroughly score 24 hours of EEG/EMG recordings, untrained staff may spend up to 6 hours to score the same amount of data. This tedious process poses not only a challenge to human scorers who often face the problem of decreasing scoring accurracies as concentration fades during such monotonous tasks. It also poses a challenge for studies to achieve a suitable statistics (with a reasonable number of mice scored) and to establish long-term studies in which potentially several days and weeks of sleep recordings need to be scored. 

Early approaches towards automating the sleep scoring process date back to the late 1960s~\cite{Drane1969}. Since then, the field has made rapid progress over the years (see~\cite{Sunagawa2013} for a historic overview), yielding methods that aim at interactively supporting human scorers or even fully automating the scoring process. Most methods rely on hand-crafted features (feature engineering), many of which are derived from power spectra of the EEG (see, e.g., Refs.~\cite{Benington1994,Kohtoh2008,Brankack2010}) where researchers focused on classical frequency bands in rodents (delta, theta, and sigma bands). Other approaches include features derived from bispectra \cite{Swarnkar2010}, or wavelet coefficients (see, e.g., Ref.~\cite{Ebrahimi2008}). While feature engineering is often the only feasible way to create predictive models in cases in which data is not abundant, this process relies on expert knowledge and may not yield features that are optimal for the sleep scoring task at hand. However, if data is abundant, feature engineering can in many cases be replaced by automated feature learning, which can lead to better prediction results as has been shown in the context of deep learning in various disciplines in recent years~\cite{Goodfellow2016}.

\begin{figure*}
 \centering
 \includegraphics[width=15cm,keepaspectratio=true]{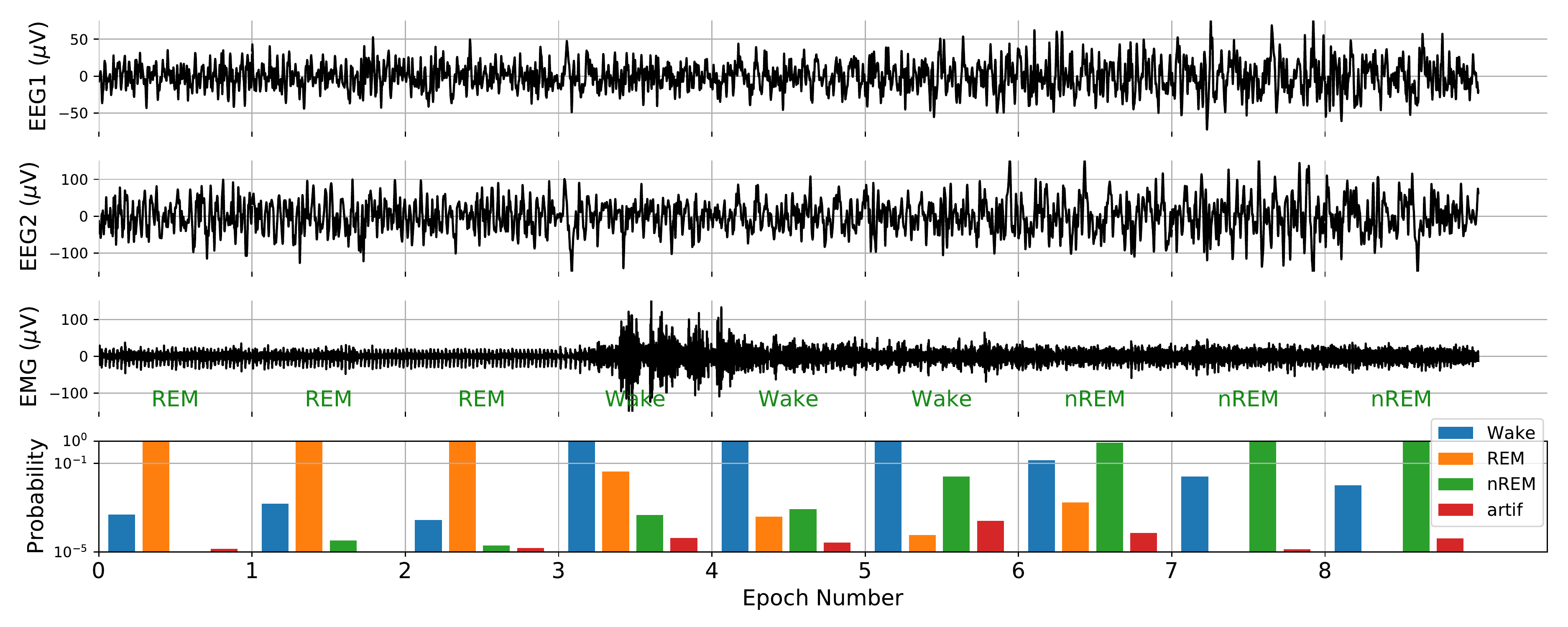}
  \caption{\textbf{Automated sleep scoring.} The model predicts from up to three input channels (EEG, top two panels; EMG, third panel) the probabilities of an epoch to belong to different sleep classes (third panel) and the probability of the input data to contain artifacts. The sleep class associated with the maximum probability is outputted as the prediction of the model. Each epoch has a duration of 4\,s; ground truth labels as determined by DS are shown in green letters in the third panel.}
 \label{fig:example_predictions}
\end{figure*}

We introduce a deep neural network model that is trained end-to-end to predict different states of consciousness in mice from EEG and EMG recordings (see Fig.~\ref{fig:example_predictions}). Instead of engineering features, features are learnt automatically by the model. The model architecture (see Fig.~\ref{fig:model_architecture}) and the training of the model draws on recent advances in convolutional neural network research which has demonstrated the usefulness of learnt features for time sequenced data in a number of applications (e.g., generative modeling of audio data~\cite{Oord2016}, machine translation~\cite{Gehring2017}). Our model achieves excellent scoring results for out-of-sample data and in artifact-free conditions. Furthermore, we demonstrate that increasing the number of recording channels can partially counteract the decrease of scoring performance for artifact-contaminated data.

\section{Methods}
\subsection{Data}
\label{sec:data}
The data was recorded from 22 mice (age 10-11 weeks, male, C57BL/6 strain). All mice were kept at the animal facility of the Center for Regenerative Therapies of the Technische Universit\"at Dresden, Germany. All applicable local and federal regulations of animal welfare in research were followed. The experiments were approved by the responsible authority, Landesdirektion Sachsen, Germany. The mice were chronically implanted with a synchronous recording system of 1 EMG and 2 EEG electrodes (8201, Pinnacle Technology Inc., Lawrence, KS) 10-14 days before the first recording. Four stainless steel screws sitting on top of the cortex at the following coordinates (relative to bregma) were used as EEG electrodes: anterior-posterior (AP): +2\,mm, medial-lateral (ML): +1.5\,mm (EEG2) / ML: -1.5\,mm (ground) and AP: -4\,mm, ML: +1.5\,mm (EEG1) / ML: -1.5\,mm (reference). With this setup, we sampled the electrical activity of the frontal (EEG2) and the parieto-occipital lobes (EEG1), respectively. Two flexible stainless steel wires were inserted into the neck muscles to measure the EMG signal. All electroencephalographic (EEG) and electromyographic (EMG) time series were sampled at 400\,Hz. 82 recordings, each lasting 24 hours and containing all three channels, were available for the creation of training, validation and test sets.

All recordings were divided into non-overlapping consecutive \emph{epochs}
of 4 seconds.  One of the authors (DS) scored each epoch manually as sleep stage ``Non REM'', ``REM''
or ``Wake''.
In addition, each epoch was scored according to whether recording
artifacts were present (artifact-contaminated epoch) or not
(artifact-free epoch).  Manual scoring was performed using our own
customized software \emph{edfView} \cite{Schwabedal17a} that was configured to show $5$ consecutive epochs of all
three channels, and to allow our scoring specialist to assign one of the labels to the middle epoch: The Wake state is characterized by high EMG activity (active locomotion, higher muscle tone even when resting) and a mixed frequency EEG signal (without a clear peak at a certain frequency in the power spectrum). Non REM sleep shows lower EMG activity compared to the Wake state as well as a power spectrum peak in the delta band (0.5--4\,Hz). REM sleep is usually characterized by very low EMG activity (with only seldom muscle twiches and a clear power spectrum peak in the theta band (at 7--8\,Hz)).

Time series were low-pass filtered with a 4th-order Butterworth filter (cutoff frequency: 25.6 Hz) using a forward-backward scheme to ensure zero-phase filtering \cite{Smith2002}. The filtered time series were downsampled from 400~Hz to 64~Hz by linearly interpolating between neighboring sampling points. Each of the 82 recordings was divided into different parts, and segments (epochs) of these parts were assigned to the training, test, and validation set, respectively (cf. Tab.~\ref{tab:1}). We made sure that data of the test set came from different mice. 

Table~\ref{tab:2} summarizes the number of labeled epochs for the whole dataset. The most frequent class was ``Wake'' ($57\%$), followed by ``Non REM'' ($33\%$) and ``REM'' ($5.9\%$) sleep. Artifacts were most pronounced during Wake ($3.6\%$) while they were barely present during REM sleep ($0.04\%$). Such large class imbalances are known to negatively affect training results. The large variability in the frequencies of class members can adversely affect the model fitting process. We accounted for the class imbalance by resampling with replacement and used the resulting \emph{rebalanced training set} for training our models.  Indeed, we observed better fitting results with this rebalanced training set, and the sampling probabilities (class percentages) used to produce the rebalanced training set are reported in table~\ref{tab:2}.

\begin{table}
\caption{\textbf{Characteristics of the data set.}}
\begin{tabular}{r|rrrrr}
 & training set & test set & validation set & all data\\\hline
 recording days & 65 & 4 &13 & 82\\
 number of epochs & 1,407,706 & 84,694 & 275,599 & 1,767,999\\
 percentage & 79.6\,\% & 4.8\,\% & 15.6\,\% & 100\,\%
\end{tabular}
 \label{tab:1}
\end{table}

\begin{table}
 \caption{\textbf{Statistics of labeled epochs} of the whole dataset (``\emph{all data}''), of the training set and the rebalanced training set (see text). NR: ``Non REM sleep'', R: ``REM sleep'', W: ``Wakefulness''. (A): artifact-contaminated. perc: percentage, (reb.) train. set: (rebalanced) training set, \# epochs: number of epochs.}
\tabcolsep=0.11cm
\begin{tabular}{r|rrrrrr}
 label & NR & NR (A) & R & R (A) & W & W (A)\\\hhline{=|======}
 \emph{all data} & & & & & &\\
 \# epochs & 586,832 & 1,973 & 103,546 & 672 & 1,010,412 & 64,564\\
 perc. & 33.19\,\% & 0.11\,\% & 5.86\,\% & 0.04\,\% & 57.15\,\% & 3.65\,\%\\\hhline{=|======}
 \emph{train. set} & & & & & &\\
  \# epochs & 465,134 & 1,290 & 82,646 & 429 & 807,419 & 50,788\\
 perc. & 33.04\,\% & 0.09\,\% & 5.87\,\% & 0.03\,\% & 57.36\,\% & 3.61\,\% \\\hhline{=|======}
 \emph{reb. train. set} & & & & & &\\
 \# epochs & 305,061 & 165,145 & 188,307 & 164,293 & 406,125 &      178,775 \\
 perc. & 21.67\,\% & 11.73\,\% & 13.38\,\% & 11.67\,\% & 28.85\,\% & 12.70\,\%
\end{tabular}
 \label{tab:2}
\end{table}

\subsection{Machine-Learning Model architecture}

To closely imitate the scoring procedure (cf. section~\ref{sec:data}), our learning model uses information of $5$ consecutive time series epochs in order to classify the middle epoch to belong to one of three distinct sleep state classes and to decide whether the middle epoch contains artifacts or not.
More formally, our learning model is a function $F$ that maps $5$ consecutive epochs of multivariate time series $\vec{x}_i$, $i\in\{1,\ldots,5\}$, to the vector \pstate and to a scalar \partif. Values of \partif larger than a threshold value of $0.5$ indicate the middle epoch $\vec{x}_{3}$ to contain artifacts. The entries of \pstate are interpreted as probabilities of the middle epoch $\vec{x}_{3}$ to reflect wakefulness (Wake), REM, or non-REM sleep. The model identifies the maximum component of \pstate and outputs its associated class, thereby producing a label of the most probable sleep state.

\begin{figure*}
 \centering
 \includegraphics[width=15cm,keepaspectratio=true]{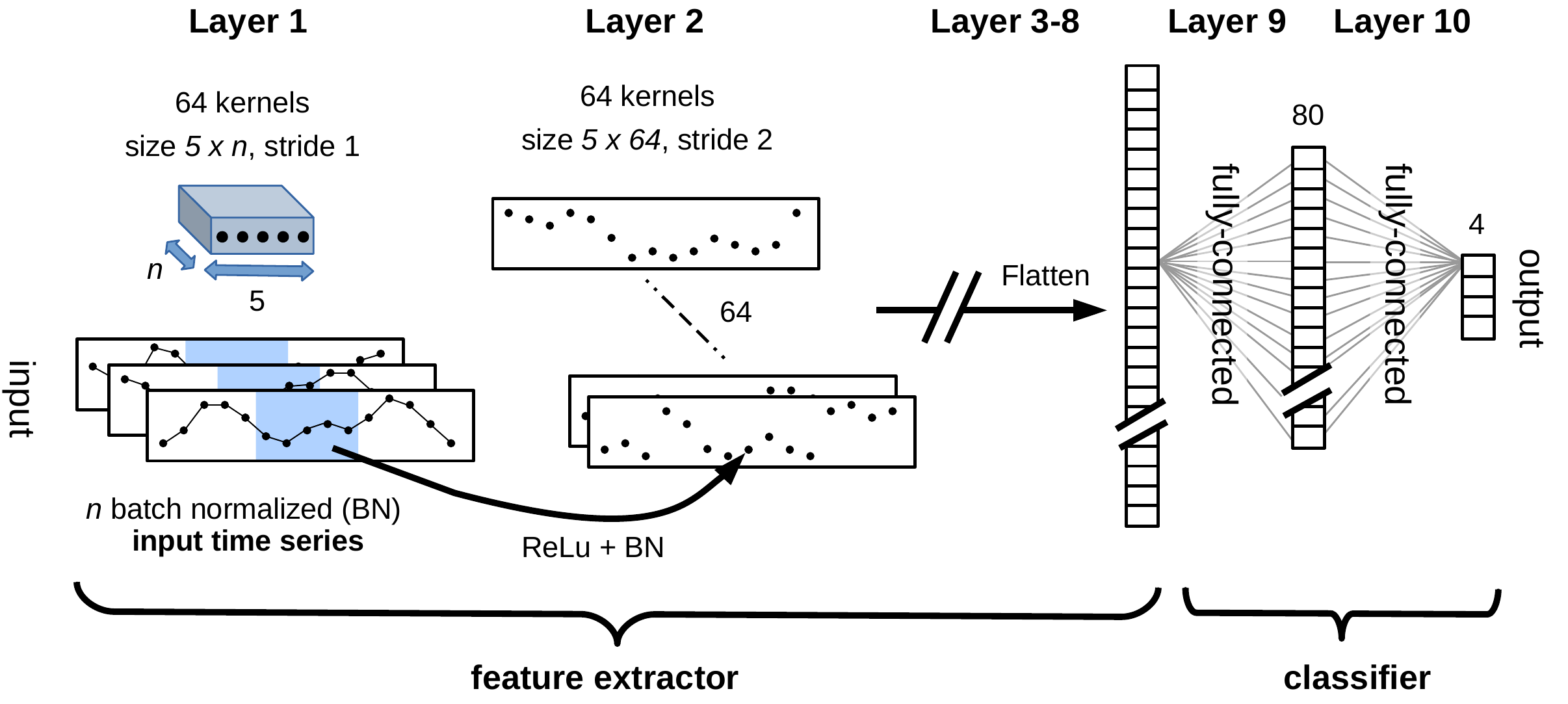}
 \caption{\textbf{Deep neuronal network model} that predicts from time series input the corresponding sleep stage class. The feature extractor consists of a hierarchy of 8 convolutional layers and produces a feature vector. The classifier is composed of 2 fully connected layers and transforms the feature vector to class probabilities as well as a probability that artifacts are present in the input time series. Features are learnt automatically during training.}
 \label{fig:model_architecture}
\end{figure*}

The model architecture (cf. fig.~\ref{fig:model_architecture}) is inspired by recent advances in deep learning \citep{LeCun2015,Oord2016} and consists of two parts: (1) A feature extractor successively downsamples and nonlinearly transforms multivariate time series to create a set of features that are used by (2) the classifier to classify the time series. The feature extractor consists of 8 convolutional layers without padding where each layer has 64 kernels of size $5$. Kernels (also called filters) are shifted with a stride of 1 in layers 1, 3, 5, and 7. The other convolutional layers use a stride of 2 and thus successively downsample the nonlinearly transformed data. We used Rectified Linear Units (ReLUs) as nonlinearities after each convolutional layer since ReLUs were observed to show superior training behavior in previous studies \citep{Nair2010,Krizhevsky2012}. To stabilize training, we reduced internal covariate shift by applying Batch Normalization (BN) \citep{Ioffe2015} after the nonlinearities of each convolutional layer\footnote{We note that in Ref.~ \cite{Ioffe2015} Batch Normalization (BN) was carried out before the nonlinearities of a layer. However, we observed our model to fit our data well when applying BN after the nonlinearities.}. In addition, we also applied Batch Normalization to the input data (i.e., before the first convolutional layer). The batch normalized 64 feature maps after the last convolutional layer are flattened and concatenated to form a single feature vector which is the output of the feature extractor.

The classifier consists of two fully connected (fc) layers, fc1 and fc2. fc1 is composed of 80 neurons with ReLU nonlinearities and processes the feature vector produced by the feature extractor. fc2 consists of 4 neurons which process the output activities of fc1.  Instead of ReLU nonlinearities, we apply a softmax on the output of the first three neurons, yielding the probabilities $p_j$, $j\in\{1,\ldots,3\}$ of the input epoch to belong to the three classes. The output of the fourth neuron is transformed into a value $p_a$ by a sigmoid function, which maps values into the interval $[0,1]$. We interpret the first three components of the resulting output vector as class probabilities while the last component shall indicate whether the input epoch contains artifacts (1) or not (0). Since fully connected layers contain many free parameters that make overfitting more likely, we employed dropout regularization (dropout probability: $0.2$) \citep{Srivastava2014} on the weights connecting the feature vector to the neurons of fc1 and on the weights connecting the output of fc1 to the neurons of fc2.

We implement the training objective by a loss function $\mathscr{L}=L_1+L_2$ that becomes small ($L_1$) if the classifier assigns a large probability to the correct class and becomes small ($L_2$) if the classifier correctly indicates whether the input contains artifacts or not. The loss is computed over a minibatch of the input. Let $c_k\in\{1, 2, 3\}$ denote the correct class index for the middle input epoch $\vec{x}_k$ and let $p_{k,c_k}$ be the predicted class probability of $\vec{x}_k$ for the correct class $c_k$. We define $L_1$ as a negative log-likelihood,
\begin{equation}
 L_1 = - \frac{1}{N_b} \sum_{k=1}^{N_b} \log(p_{k,c_k})
\end{equation}
where $N_b$ is the minibatch size. The second term $L_2$ is defined as the binary cross entropy between the predicted value $p_{k,a}$ for an input $\vec{x}_k$ and the true label $t_{k,a}\in\{0,1\}$ indicating whether an artifact is present (1) or not (0),
\begin{equation}
 L_2 = \frac{1}{N_b} \sum_{k=1}^{N_b} \left(-t_{k,a}\log{p_{k,a}} + (1-t_{k,a})\log{(1-p_{k,a})}\right).
\end{equation}
The model is differentiable and can be trained via gradient descent.

\subsection{Training}
\label{sec:training}
The model was trained by RMSprop (Root Mean Square Propagation), a variant of stochastic gradient descent where the learning rate is adapted separately for each free parameter (weight) of the model \citep{Tieleman2012}. Let $w(t)$ denote a weight and let $\Delta(t) = \frac{\partial \mathscr{L}}{\partial w}$ denote the gradient of the loss function with respect to the weight obtained for the minibatch at step $t$ of the training. We update the weight by
\begin{equation}
 w(t+1) = w(t) - \frac{\eta}{\sqrt{R(t)}} \Delta(t)
\end{equation}
where $\eta$ is the learning rate. $R$ is a running mean of the squared magnitudes of recent gradients for that weight,
\begin{equation}
 R(t+1) = \alpha R(t) + (1-\alpha) \Delta(t)^2,
\end{equation}
where we set the smoothing factor $\alpha$ to $\alpha=0.99$. After the gradients for all weights are determined, we checked whether a gradient exceeded the value $0.1$. In such a case, we rescaled all gradients such that the largest value was $0.1$ (maximum-norm normalization of gradients).

We adjusted the learning rate according to the following learning protocol: during model exploration (section~\ref{sec:model_exploration}), the learning rate $\eta$ was linearly increased from $0$ to $\eta^*=0.00128 N_b$ (warm-up period) in the first $5$ training cycles, where $N_b$ denotes the size of the minibatches. This linear scaling of the learning rate with minibatch size combined with a warm-up period was observed to lead to better generalization properties of models when trained with large minibatch sizes~\cite{Goyal2017}. A single training cycle was completed when all minibatches of the training set were used once for a stochastic gradient descent step. During the following $5$ training cycles, the learning rate was linearly decreased again to $0$ (cool-down period). For the training of the final models (section~\ref{sec:model_performance}), we linearly increased the learning rate from $0$ to $\eta^*$ in the first 5 training cycles. During the subsequent 10 training cycles, the learning rate was $\eta^*$, while during the next 5 training cycles we linearly decreased $\eta$ again from $\eta^*$ to $0$. All trainings were carried out with a minibatch size of $N_b=256$, and we did not observe moderate changes in minibatch sizes to affect training outcomes.

\subsection{Evaluation.}

To evaluate the prediction performance of our model, we employed classical techniques to characterize the quality of classifications in  binary as well as in multiclass classification settings.

\paragraph{Artifact classification} We used the \emph{$F_1$ score} (also known as \emph{$F$ measure}, \cite{Manning2008}) to quantify the prediction performance of our model to distinguish between artifact-free and artifact-contaminated epochs (binary classification) in a given sleep stage. The \fscore is a summary statistics which takes on values between $0$ and $1$, where larger values indicate better predictive performance. It is defined as the harmonic mean of \emph{precision} (\precision) and \emph{recall} (\recall),
\begin{equation}
 F_1 = \left( \frac{1}{\precision} + \frac{1}{\recall}\right)^{-1},
\end{equation}
where precision is also known as \emph{positive predictive value}. In a binary classification setting, the precision is the number of epochs for which a class was correctly predicted by the model (number of true positives) divided by the total number of epochs that were predicted by the model to belong to that class,
\begin{equation}
 \precision = \frac{\text{number of true positives}}{\text{number of true and false positives }}.
\end{equation}
The recall, also known as \emph{sensitivity}, is the number of epochs for which a class was correctly predicted by the model divided by the total number of epochs of that class,
\begin{equation}
 \recall = \frac{\text{number of true positives}}{\text{number of true positives and false negatives}}.
\end{equation}
Both, precision and recall, take on values between $0$ and $1$. 

\begin{table}
\caption{\textbf{Model exploration.} $\bar{F}_1$ (top two rows) and $F_1$ (last three rows) scores obtained on training and validation sets. l: layers, k: kernels. Light grey: best \bfscores obtained on the validation set; dark gray: slight cases of overfitting.}
\tabcolsep=0.11cm
\begin{tabular}{c|r|c|r|r|r|r|r|r|} 
 \multirow{2}{*}{\makecell{prediction \\ target}} & & & \multicolumn{3}{c|}{96 k} & \multicolumn{3}{c|}{$4$ l} \\
 & condition & set & 4 l & 8 l & 10 l & 32 k & 64 k & 96 k\\\hhline{=========}
 \multirow{4}{*}{\makecell{sleep \\ stages}}
 &no artifact  & training   & 0.95 & 0.98 & 0.98 & 0.92 & 0.94 & 0.95 \\
 &             & validation & 0.93 & \cellcolor[gray]{0.90} 0.95 & \cellcolor[gray]{0.90} 0.95 & 0.91 & 0.93 & 0.93 \\\hhline{~--------}
 &artifact     & training   & 0.98 & 1.00 & 1.00 & 0.93 & 0.97 & 0.98 \\
 &	       & validation & \cellcolor[gray]{0.8} 0.88 & \cellcolor[gray]{0.8} 0.85 & \cellcolor[gray]{0.8} 0.82 & 0.85 & 0.88 & 0.88 \\\hhline{=========}
 \multirow{6}{*}{\makecell{artifact \\ yes/no}} 
 & Non-REM     & training & 0.94 & 0.94 & 0.96 & 0.70 & 0.88 & 0.94\\
       &       & validation & 0.60 & 0.80 & 0.81 & 0.48 & 0.56 & 0.81\\\hhline{~--------}
    & REM      & training   & 1.00 & 1.00 & 1.00 & 0.84 & 0.97 & 0.99\\
        &      & validation & 0.60 & 0.76 & 0.76 & \cellcolor[gray]{0.8} 0.53 & \cellcolor[gray]{0.8} 0.63 & \cellcolor[gray]{0.8} 0.58\\\hhline{~--------}
 & Wake & training   & 0.89 & 0.95 & 0.96 & 0.82 & 0.87 & 0.89\\
&	       & validation & 0.85 & 0.86 & 0.87 & 0.81 & 0.83 & 0.84\\\hhline{---------}
\end{tabular} 
 \label{tab:model_exploration}
\end{table}

\paragraph{Sleep-stage classification} We quantified the prediction performance of our model to distinguish between different sleep stages (multiclass classification) by calculating averages of \fscores. More precisely, we converted the multiclass classification problem into multiple binary classifications: We determined the \fscore for each class $c$ (sleep stage) separately ($F_1^c$) where precision and recall were calculated with respect to the distinction between class $c$ and everything else ($\neg c$). The resulting $F_1$ scores were then averaged to arrive at a final score,
\begin{equation}
 \bar{F}_1 = \frac{1}{N_c} \sum_{c=1}^{N_c} F_1^c,
\end{equation}
where $N_c$ denotes the number of classes ($N_c = 3$ sleep stages). The \bfscore varies between $0$ and $1$ where large values indicate better predictive performance. Next to the \bfscore, we also determined confusion matrices in order to investigate whether our models systematically confused one class with another.

\section{Results}
We investigated our learning model with respect to its ability (i) to accurately predict sleep stages and (ii) to detect artifacts. In a first series of experiments, we investigated to what extent changes of our basic model architecture (as shown in figure~\ref{fig:model_architecture}) affected prediction performance. In this \emph{model exploration} step, we evaluated the performance of different model architectures on the validation set (see section~\ref{sec:model_exploration}). Based on these results, we chose a final model architecture. In a second series of experiments (see section~\ref{sec:model_performance}), we addressed the question whether the predictive performance changes when varying the number of input channels presented to the model. Here we evaluated the performance of our model on the test set.

\subsection{Model exploration}
\label{sec:model_exploration}

We varied our model architecture (cf.~figure~\ref{fig:model_architecture}) and investigated which of the resulting architectures yielded the best prediction performance. We varied (i) the number of convolutional layers and (ii) the number of kernels per convolutional layer. The input for all models consisted of three input time series (EEG1, EEG2, EMG) that were simultaneously recorded, and each model was trained as described in section~\ref{sec:training}.

When increasing the number of layers or the number of kernels, we increase the complexity of the model and thus its ability to fit the training data. Indeed, we observed \fscores obtained on the training set (see table~\ref{tab:model_exploration}) to increase with an increasing number of layers and number of kernels per layer. Best \bfscores were obtained for the architecture with largest model complexity (10 convolutional layers with 96 kernels each) which reach $0.98$ and $1.00$ for sleep stage classification from artifact-free and artifact-contaminated epochs, respectively. For the same model architecture in the artifact-free condition, our model generalized well to unseen epochs as indicated by an \bfscore of $0.95$ for the validation set. Due to this high value, we expect our model to very reliably predict sleep stages on unseen and artifact-free epochs.

Our models were able to fit artifact-contaminated time series in the training set, reaching \bfscores larger than $0.92$ for all model architectures (cf.~table~\ref{tab:model_exploration}). However, as compared to the artifact-free case, our models were not able to generalize as well, which was indicated by lower \bfscores between $0.82$ and $0.88$ on the validation set. This phenomenon may be related to class imbalances of the training set, where artifact-free epochs were much more frequent than artifact-contaminated epochs (cf.~table~\ref{tab:2}). We observed the same phenomenon, namely less generalization performance, for the prediction of artifacts under the condition of non-REM, REM, and Wake stages (cf.~table~\ref{tab:model_exploration}, last three rows).

We noticed tendencies of slight overfitting (decreasing \bfscores for increasing model complexity for the validation set) when predicting sleep stages from artifact-contaminated epochs and increasing the number of layers (see dark gray cells in table~\ref{tab:model_exploration}). This tendency was also observed for the prediction of artifacts from epochs recorded during REM sleep and increasing the number of kernels. While we do not consider these cases of overfitting to be severe, we chose our final model architecture, $8$ layers and $96$ kernels, as a compromise between high \bfscores for sleep stage prediction (see light gray cells in table~\ref{tab:model_exploration}) and reduced overfitting.

\begin{table}[t]
\begin{threeparttable}
\caption{\textbf{Model performance.} $\bar{F}_1$ (top two rows) and $F_1$ (last three rows); ic: input channels; grey cells: largest $\bar{F}_1$ scores for predicting sleep stages in artifact-free and artifact-contaminated epochs in the test set.\tnote{1}}
\tabcolsep=8.0pt
\begin{tabular}{c|r|c|r|r|r|} 
 \multirow{2}{*}{\makecell{prediction \\ target}} & & & \multicolumn{3}{c|}{96 kernels, 8 layers}\\
 & condition & set & 1 ic & 2 ic & 3 ic\\\hhline{======}
 \multirow{6}{*}{\makecell{sleep \\ stages}}
 &no artifact  & training   & 0.95 & 0.97 & 0.98 \\
 &             & testing    & 0.93 & 0.95 & \cellcolor[gray]{0.8} 0.96 \\
 &             & validation & 0.92 & 0.94 & 0.95 \\\hhline{~-----}
 &artifact     & training   & 1.00 & 1.00 & 1.00 \\
 &             & testing    & 0.82 & 0.92 & \cellcolor[gray]{0.8} 0.93 \\
  &	           & validation & 0.77 & 0.85 & 0.84  \\\hhline{======}
 \multirow{9}{*}{\makecell{artifact \\ yes/no}} 
 & Non-REM     & training   & 0.87 & 0.91 & 0.95 \\
&              & testing    & 0.12 & 0.15 & 0.17 \\
 &             & validation & 0.68 & 0.68 & 0.81\\\hhline{~-----}
    & REM      & training   & 0.99 & 1.00 & 1.00\\
 &             & testing    & 0.40 & 0.36 & 0.56 \\
        &      & validation & 0.68 & 0.65 & 0.73 \\\hhline{~-----}
 & Wake & training   & 0.92 & 0.94 & 0.96 \\
 &             & testing    & 0.74 & 0.75 & 0.83 \\
 &	           & validation & 0.79 & 0.80 & 0.87 \\\hhline{------}
\end{tabular} 
 \label{tab:model_performance}
 \begin{tablenotes}
  \item[1]  Small differences between values reported in table~\ref{tab:model_exploration} and in table~\ref{tab:model_performance} are due to a retraining of the models and the stochastic nature of the optimization algorithm (stochastic gradient descent).
 \end{tablenotes}
 \end{threeparttable}
\end{table}

\subsection{Model performance}
\label{sec:model_performance}

We studied the predictive performance of the final model and whether and how it changed when predictions were based on a different number of simultaneously recorded time series (input channels). This question becomes important when experimental restrictions and setups do not allow for all three channels to be recorded. A reliable prediction of sleep stages and artifacts based on few channels is highly desirable in such cases.

\begin{figure*}[t]
 \centering
 \includegraphics[width=15cm,keepaspectratio=true]{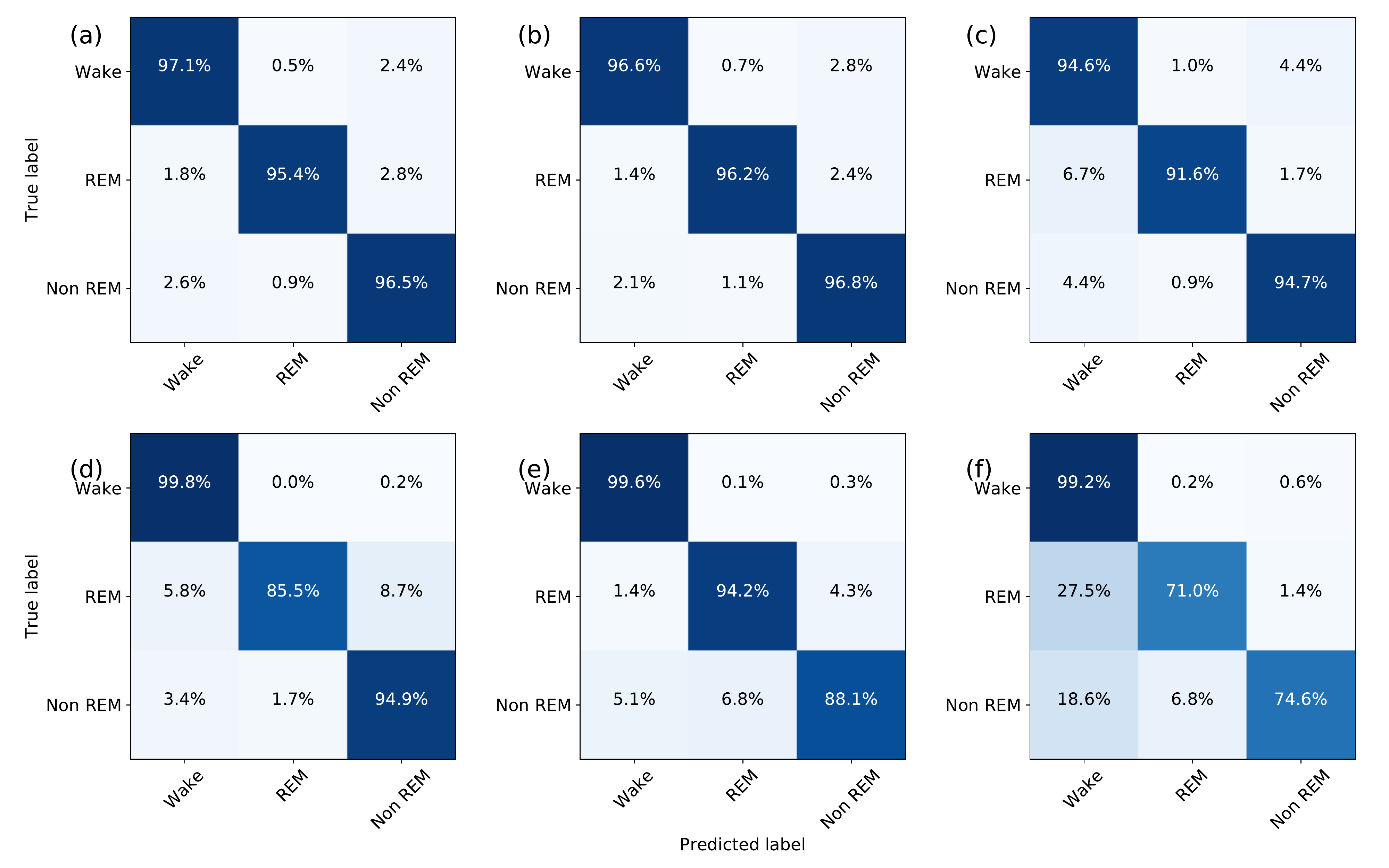}
 \caption{Confusion matrices for predicting sleep stages with the final model (8 layers, 96 convolutional kernels) from artifact-free epochs (a--c) and artifact-contaminated epochs (d--f) obtained on the test set. The final model predicted sleep stages from three input time series (a, d), two input time series (b, e), and one input time series (c, f). Matrices show the percentages of epochs that were correctly or incorrectly assigned to a sleep stage by our model.}
 \label{fig:confusion-matrices}
\end{figure*}

We varied the number of input channels from one channel (``EEG2''), to two (``EEG2, EMG''), and three (``EEG1, EEG2, EMG'') and trained one final model for each of the three cases using the final architecture (i.e., 8 convolutional layers with 96 kernels each) as determined in the previous section. \bfscores for sleep stage prediction and \fscores for artifact prediction obtained for all datasets are shown in table~\ref{tab:model_performance}.
We observed \bfscores and \fscores obtained on the training set to increase with increasing the number of input channels, indicating our models to better predict sleep stages or artifacts the more input channels were available.
This tendency was also observed for the test set and, as such, for epochs not used for training. We obtained best \bfscores for sleep stage prediction of $0.96$ and $0.93$ (grey cells in table~\ref{tab:model_performance}) for the three channels condition and for artifact-free and artifact-contaminated epochs, respectively. More importantly, \bfscores obtained on the test set only slightly decreased when decreasing the number of input channels. This indicates that our model is able to reliably predict sleep stages even in experimental setups in which only one channel is recorded.

\fscores for predicting artifacts obtained low values on the test set, e.g. $0.12$ for one input channel and artifact prediction in Non-REM epochs. We hypothesized that these low values are related to the small size of the test set that only encompasses $4.8\,\%$ of the labeled total data (cf. table~\ref{tab:1}). Since the test set is small, it has only few artifact-contaminated epochs that are, moreover, assigned to different sleep stages. Due to this difficult statistical situation, we also reported \fscores (cf.~table~\ref{tab:model_performance}) obtained on the validation set (which was larger than the test). Indeed, for the validation set and the prediction of artifacts, we observed \fscores which are close to those reported in the previous section. 

To offer insight into the prediction of individual labels, we provide confusion matrices (figure~\ref{fig:confusion-matrices}) grouped for artifact-contaminated and artifact-free epochs. 
For three input channels and for artifact-free epochs, our final model correctly classified 95.4\,\% of all ``REM'' epochs, 96.5\,\% of all ``Non REM'' epochs, and $97.1\,\%$ of all ``Wake'' epochs. These numbers only slightly decrease when reducing the number of input channels, where we obtained $91.6\,\%$ as the lowest fraction of correct predictions for artifact-free REM epochs (cf. figure~\ref{fig:confusion-matrices}). 
As expected, prediction performance worsened for artifact-contaminated epochs, where our model showed tendencies to confuse ``Non REM'' or ``REM'' sleep stages with ``Wake''. This confusion was most pronounced when using a single input channel ($27.5\,\%$ of ``REM'' epochs were misclassified as ``Wake''). Increasing the number of input channels, however, led to more robust predictions in artifact-contaminated epochs, i.e. to a larger fraction of correctly predicted sleep stages.

\section{Discussion}
We introduced a deep neural network model that predicts different states of consciousness in mice from EEG and EMG recordings. Unlike many previous approaches towards automated sleep scoring, our model does not rely on manually defined features (such as power in frequency bands). Instead, our model was trained end-to-end, thereby automatically learning features of the data that were successfully used by the classifier of the model to distinguish between different sleep stages. While many previous approaches predicted sleep stages on epochs of 10 seconds length, our model allows for higher time resolution and was able to make reliable predictions for epochs of 4 seconds length. Such time resolutions are important to capture short arousals and frequent changes in sleep states that are typical in mice. High time resolutions particularly call for approaches towards automatizing potentially time-consuming manual scoring procedures. 

We observed our model to achieve high specificity and sensitivity (as indicated by large \fscores) on artifact-free out-of-sample (test) data. Decreasing the number of input channels that were available for our model to infer sleep stages led only to a slight decrease of prediction performance for artifact-free data. This indicates that our model will yield good prediction performance even for experiments in which constraints prohibited to simultaneously capture multiple recording modalities. For artifact-contaminated data, \fscores became better when the model based its predictions on more input channels. This observation agreed with our expectation that recordings will likely contain partially redundant information which renders inferring sleep classes from artifact-contaminating epochs easier if other recording channels are available. Nevertheless, prediction performance of our model was decreased for artifact-contaminated data when compared to artifact-free data. We speculate that the deteriorating effect of artifacts may also be related to our training set that contained much more artifact-free than artifact-contaminated epochs. We addressed these class imbalances by sampling with replacement but expect larger training sets to allow future models to even further increase their prediction performance on artifact-contaminated data.

In our study, we assessed prediction performance on test data which was created from 22, male, 10-11 week old, genetically identical C57BL/6 mice with a fixed electrode placement scheme. This restriction was due to experimental constraints when creating the data set and limits our ability to draw conclusions on prediction performance for other mice strains and electrode placements. However, we are confident that our deep learning model can be successfully employed to predict sleep stages in other mice strains, models and/or other implantation schemes after retraining. Such a retraining could substantially profit from transfer learning techniques~\cite{Pan2010} which have the potential to significally reduce the amount of training data required. Next to the aforementioned restriction, we also did not assess inter-rater and intrarater variability, i.e. the scoring variability between different persons scoring our data and the same person scoring the same data multiple times, respectively. Such assessments become increasingly challenging as the amount of data to score increases. We speculate, however, that intrarater variability sets an upper limit to the maximum achievable sensitivity and specificity of our model. Also, previous studies observed inter-rater reliability to be never 100\,\%  but rather in the range of 83--96\,\%~\cite{Kreuzer2015}. Future studies on model architectures like the one introduced here that assess rater variability will be able to relate scoring accuracies achieved by models to those achieved by human scorers.

We expect models like ours (a TensorFlow.js implementation is available online  \cite{Schwabedal2018}) to facilitate or even enable long-term studies in sleep research. Automating the time-consuming process of sleep scoring will contribute to support sleep research as a data-driven field and will help researchers and lab personnel to become more productive. This is in line with what we expect to happen in the early 21st century: The automation of cognitive repetitive tasks. In this context, deep learning models and automated feature learning will play an important role.




\end{document}